
\headline={\ifnum\pageno=1\firstheadline\else
\ifodd\pageno\rightheadline \else\leftheadline\fi\fi}
\def\firstheadline{\hfil}
\def\rightheadline{\hfil}
\def\leftheadline{\hfil}
        \footline={\ifnum\pageno=1\firstfootline\else\otherfootline\fi}
\def\firstfootline{\rm\hss\folio\hss}
\def\otherfootline{\hfil}

\font\tenrm=cmr10

\font\elevenbf=cmbx10 scaled\magstep 1
\font\elevenrm=cmr10 scaled\magstep 1
\font\elevenit=cmti10 scaled\magstep 1

\font\ninerm=cmr9

\nopagenumbers
\hsize=6.0truein
\vsize=8.5truein
\parindent=1.5pc
\baselineskip=10pt
\centerline{\elevenbf SUSY CONTRIBUTIONS TO THE $Z\rightarrow
3\gamma$\
DECAY\footnote*{{\ninerm\baselineskip=11pt
Talk presented at DPF-94, U. of New Mexico, Albuquerque,
New Mexico, USA, August 2-6, UQAM-PHE-94/11.}}}
\vglue.5cm
\centerline{\elevenrm HEINZ K\"ONIG}
\baselineskip=12pt
\centerline{\elevenit D\'epartement de Physique}
\baselineskip=12pt
\centerline{\elevenit Universit\'e du Qu\'ebec \`a Montr\'eal}
\baselineskip=12pt
\centerline{\elevenit C.P. 8888, Succ. Centre Ville, Montr\'eal}
\baselineskip=12pt
\centerline{\elevenit Qu\'ebec, Canada H3C 3P8}
\vglue 0.5cm
\centerline{\tenrm ABSTRACT}
\vglue 0.5cm
  {\rightskip=3pc
 \leftskip=3pc
 \tenrm\baselineskip=12pt
 \noindent
In this talk I comment on the contributions of all
three types of particles (scalars, fermions and
bosons) to the three--photon Z decay within the
standard model (SM) and its extensions
to the 2 Higgs doublet model (2 HDM) and the
minimal supersymmetric extension of the standard
model (MSSM).
I also comment briefly on the charginos
contribution to this decay rate. Finally
I comment on the fermions and scalars contributions to
the three--gluon Z decay.
\vglue 0.5cm}\noindent
{\elevenbf 1. Fermionic contribution}
\vglue 0.5cm
\baselineskip=12pt
\elevenrm
The fermionic contribution was considered a while ago
in [1--4] and complete analytical results were presented
in [5]. There is only one type of diagram with four
internal fermions within the loop. The divergencies
as well as the $\gamma_5$\ terms drop out when summed
over all six possible permutations of the photons (or, what
is topologically equivalent, when summed over diagrams
where the fermions circulate in one direction and in
the opposite direction). The result is
written as:
$$\eqalignno{\Gamma_F(Z\rightarrow 3\gamma)=&{{\alpha^4}
\over{\sin^2\Theta_W\cos^2\Theta_W}}\bigl (3\sum_q e_q^3 V_q+
\sum_L e_L^3V_L\bigr )^2{{m_Z}\over{3\cdot 3!4\pi^3}}X_F&(1)\cr
&\approx 1.05\times 10^{-9}\ {\rm GeV}\cr}$$
with $m_Z=91.1$,
$\alpha=1/128$, $\sin^2\Theta_W=0.23$\ and $X_F
\approx 15$. $V_F=(T_{3F}-2e_F\sin^2\Theta_W)/2$.
The result is given for five flavours
of light quarks since the top quark decouples
rapidily for $m_{\rm top}\ge m_Z/2$\ as was shown in
[5], where $\Gamma(Z\rightarrow 3\gamma)$\ was
plotted as a function of the top quark mass.
A small top quark mass enhances the result by
approximately 50\% [5].
\vglue 0.5cm\noindent
{\elevenbf 2. Bosonic contribution}
\vglue 0.5cm
The bosonic contribution were presented in [6--8]. Here
we have four different kind of diagrams.
Each diagram is divergent, but the sum is finite.
In [6] the authors made extensive use of the transversality
of the amplitude (that is $q_\mu M^{\mu\alpha\beta\gamma}=0=
p_1^\alpha M^{\mu\alpha\beta\gamma}=p_2^\beta
M^{\mu\alpha\beta\gamma}=p_3^\gamma M^{\mu\alpha\beta\gamma}$,
where $q$\ is the momenta of the Z boson and $p_{1,2,3}$\
the momentas of the photons and $\mu,\alpha,\beta$\ and
$\gamma$\ their Lorentz indices respectively), gauge invariance
and the masslessness of the photons, which sets many terms
in the amplitude identical to 0.
After expansion in the W boson
mass, which was kinematically valid, (as explained in [6])
the result is:
$$\eqalignno{\Gamma_W(Z\rightarrow 3\gamma)=&{{\alpha^4}
\over{\sin^2\Theta_W\cos^2\Theta_W}}
{{m_Z}\over{3\cdot 3!4\pi^3}}X_W&(2)\cr
&\approx 2.03\times 10^{-11}\ {\rm GeV}\cr}$$
with $X_W\approx 0.16$. Which gives us a number about
50 times smaller than the fermionic contribution.
\hfill\break\indent
In [5] the authors present the full analytical W bosons and
fermions contribution to the three--photon
Z decay, including the interference term
of the W bosons with the larger
fermion loop contribution, which enhances the width by
about 27\% from $1.05\times 10^{-9}$\ to $1.35\times 10^{-9}$.
Furthermore the authors show that there are strong
relations between the fermionic and bosonic amplitudes.
\vglue 0.5cm\noindent
{\elevenbf 3. Scalar contribution}
\vglue 0.5cm
Since there are no charged scalar particles in the
SM this kind of contribution only occurs in models
beyond the standard model such as the 2 HDM or the
MSSM. In [9] I present a detailed analysis. The diagrams
we have to consider are exactly the same as in the
bosonic case with the W bosons replaced by
the charged Higgs bosons. As in the bosonic case the divergencies
cancel when summed over all diagrams in a non trivial
way. In [9] I have shown through explicit calculation
that the quadratic terms $(m_{H^+}/m_Z)^2$\ are identical
to 0 after expansion with respect to the Higgs mass squared.
This is a necessary consequence of gauge invariance,
as was shown. It is also a direct consequence of the effective
gauge invariant Lagrangian for the four vector interaction. This
term has to be of the form ${\cal L}=G B_{\mu\nu}B^{\mu\nu}
W^i_{\rho\sigma}W_i^{\rho\sigma}$, where $B_{\mu\nu}$\ and
$W^i_{\rho\sigma}$\
are the tensor fields of the U(1) and SU(2) gauge fields
respectively and therefore gives G a dimension of
$1/m^4$. As was done in the bosonic case
I made use of the masslessness of the photons
and showed the transversality of the amplitude.
As a final result I have:
$$\eqalignno{\Gamma_{H^+}(Z\rightarrow 3\gamma)=&
\alpha^4\cot^2 2\Theta_W
{{m_Z}\over{3\cdot 3!4\pi^3}}X_{H^+}
\bigl ({{m_Z}\over{m_{H^+}}}\bigr )^8&(3)\cr
&\approx 3.75\times 10^{-16}
\bigl ({{m_Z}\over{m_{H^+}}}\bigr )^8\
{\rm GeV}\cr}$$
with $X_{H^+}\approx 6\times 10^{-6}$. This result
is more than a factor of $5.4\times 10^4$\ smaller than the
W boson contribution and even a factor of $2.8\times
10^6$\ smaller than the fermionic contribution.
\hfill\break\indent
In [10] a more
general analysis of the contribution of all three types
of particles to the three--photon Z decay using
supergraph techniques was presented.
The authors there obtain strong relations between
the fermionic, bosonic and scalar amplitudes.
In [11] a summary of the contribution of all three
types of particles
was given.
\vglue 0.5cm\noindent
{\elevenbf 4. MSSM contribution}
\vglue 0.5cm
In the MSSM there are besides the charged Higgs
bosons, charged scalar leptons and scalar quarks. The
calculation is exactly the same as for the
charged Higgs. We only have to replace $\cot\Theta_W$\
in Eq.(3) by $e_{q_{L,R}}(T_{3q_{L,R}}-e_{q_{L,R}}\sin^2\Theta_W)/
\sin\Theta_W\cos\Theta_W$. The summation over all generations
(including the colour factor) and the scalar partner of the left and
right handed fermions leads to
$\lbrack 3(1-88\sin^2\Theta_W/27)/\sin\Theta_W
\cos\Theta_W\rbrack^2\approx 3.19$, which is only a factor 8
larger than $\cot^2\Theta_W\approx 0.41$\ and therefore
gives us still a result far below the SM contribution.
Mixing of the scalar fermions, which
is proportional the fermion masses and therefore relevant
only for the top quark, does not change the result
to the three--photon Z decay rate
since no new couplings are introduced.
The scalar top quark contribution is only divided in
different terms suppressed by the mixing angles.
\hfill\break\indent
In the MSSM there are also
new fermions which contribute to the three--photon
 Z decay: the charginos, the
mass eigenstates of the fermionic partners of the
W bosons (winos) and the charged Higgses (Higgsinos).
The mass eigenvalues, diagonalizing angles and
couplings are well defined and in the notation
of Eq.(1) given by $V_{\tilde\chi_i}=
\lbrack 2\cos^2\Theta_W-(U^2_{i2}+V^2_{i2})/2\rbrack/2$.
U and V are the diagonalizing angles of the charginos.
With the parameters given in Eq.(1) I have $3e_L^3V_L\approx
6\times 10^{-2}, 9e_d^3V_D\approx 5.8\times 10^{-2}$\
and $9e_u^3V_u\approx 0.26$\ ($0.17$\ without the top quark).
For a wide range of the gaugino mass ($0\le m_{g_2}\le
400\ {\rm GeV}$),
the mixing parameter $\mu$\
($-500\ {\rm GeV}\le\mu\le +500\ {\rm GeV}$) and the ratio
of the Higgs vacuum expectation values ($1\le \tan\beta\le 10$)
I obtain $0.52\le e^3_{\tilde\chi_1} V_{\tilde\chi_1}\le 0.77$\ and
$0.27\le e^3_{\tilde\chi_2} V_{\tilde\chi_2}\le 0.52$.
 That is the couplings
of the charginos are even higher than those of the up quarks.
Unfortunately the couplings of the lighter chargino
eigenstate is smaller than the heavier one and from experiment
we know that the lighter chargino mass eigenstate has to be
larger than half of the Z mass and therefore decouples
like the top quark from the three--photon Z decay.
\vglue 0.5cm\noindent
{\elevenbf 5. Final remarks}
\vglue 0.5cm
As a final remark I want to mention that the results
given here can also be used for the three--gluon Z decay.
In [3] it was shown that from the three--gluon self couplings
there are also axial vector couplings contributions
to this decay rate, although much smaller than
the vector couplings. As a result they have
$\Gamma_F(Z\rightarrow 3g)\approx 6.5\times 10^{-6}$\
GeV, which is a factor of 6200 larger than the
three--photon Z decay rate. When including scalar
quarks we are lead to four different new types of
diagrams with three--gluon self couplings. The calculations
shows that when using gauge invariance, after Feynman integration
and after summation of all permutation that each diagram itself
leads to a null result. To obtain the three--gluon Z decay
from the three--photon Z decay we therefore only have
to replace $e^3\cos2\Theta_W$\ by $g_s^3{1\over 2}
d_{abc}(T_{3q_{L,R}}-e_{q_{L,R}}\sin^2\Theta_W)$\ which
leads to a factor of about 7400 larger than the scalar
quark contribution to the three--photon Z decay.
Unfortunately, experimentally this decay mode is
indistinguishable from the similiar $Z\rightarrow q
\overline q g$\ decay.
\vglue 0.5cm\noindent
{\elevenbf 6. Conclusions}
\vglue 0.5cm
We have seen that fermions lead to the largest
contribution to the three--photon Z decay within
the SM, followed by the W boson contribution,
which is about a factor of 50 smaller.
When extending the SM to the 2 HDM or the
MSSM we have to include the scalars, which
lead to a contribution several orders of magnitude
smaller than the SM contribution.
\hfill\break\indent Considering
that the experimental upper limit is
$8.1\times 10^{-6}$\ [12] and therefore even a factor
of almost $10^4$\ smaller than the highest
contribution of the SM this decay mode turns
out not to be a very good one to test neither the SM
nor the 2 HDM nor the MSSM. Decay rates
of this order can only be obtained by composite
models [13--14] or, as was shown recently, in monopole
models [15].
\vglue 0.5cm\noindent
{\elevenbf References}
\vglue 0.5cm
\item{[\ 1]}V.N. Baier, E.A. Kurayev and V.S. Fadin,
{\elevenit Sov.J.Nucl.Phys.} {\elevenbf 31}(1980)364.
\item{[\ 2]}M.L. Laurson, K.O. Mikaelin and A. Samuel,
{\elevenit Phys.Rev.}
{\elevenbf D21}(1981)2795.
\item{[\ 3]}J.J. van der Bij and E.W.N. Glover,
{\elevenit Nucl.Phys.} {\elevenbf B313}
(1989)237.
\item{[\ 4]}J.J. van der Bij and E.W.N. Glover,
{\elevenit "Rare decays"}, in Z physics
at LEP 1, CERN 89-08 Vol.2 p.30.
\item{[\ 5]}E.W.N. Glover and A.G. Morgan,
{\elevenit Z.Phys.-Particles and
Fields} {\elevenbf C60}(1993)175.
\item{[\ 6]}M. Baillargeon and F. Boudjema,
{\elevenit Phys.Lett.} {\elevenbf B272}(1991)158.
\item{[\ 7]}X.Y. Pham, {\elevenit Phys.Lett.}
{\elevenbf B272}(1991)373
\item{[\ 8]}F. Dong, X. Jiang and X. Zhou, {\elevenit Phys.Rev.}
{\elevenbf D47}(1993)214, {\elevenbf D46}(1992)5074.
\item{[\ 9]}H. K\"onig, {\elevenit Phys.Rev.} {\elevenbf D50}
(1994)602.
\item{[10]}Z. Bern and A.G. Morgan,
{\elevenit Phys.Rev.} {\elevenbf D49}(1994)6155.
\item{[11]}J. Ho\v rej\v s\'\i\ and M. St\"ohr, {\elevenit
"One--Loop induced effective on--shell $Z\gamma\gamma\gamma$\
interactions"}, PRA--HEP--94/4 to be published in {\elevenit
Z.Phys.-Particles and Fields} {\elevenbf C}.
\item{[12]} M. Sarakinos,  {\elevenit "Search for
$Z\rightarrow\gamma\gamma\gamma$\ Decays at L3"} these Proceedings.
\item{[13]}D. Treille et al.,{\elevenit "Compositeness at LEPII"},
CERN 87-08.
\item{[14]}F. Boudjema and F.M. Renard, {\elevenit "Compositeness"},
in Z physics at LEP 1, CERN 89-08 Vol.2 p.205.
\item{[15]}A. de R\'ujula, {\elevenit "Effects of Virtual
Monopoles"}, hep-th/9405191.
\vfill\break\noindent
\eject\bye